# Atomic-resolution imaging of magnetism via ptychographic phase retrieval

**Authors: Jizhe Cui**[1,2,3,†], **Haozhi Sha**[1,2,3,†], **Wenfeng Yang**[1,2,3], **and Rong Yu**[1,2,3,*]

**Affiliations:**

[1]School of Materials Science and Engineering, Tsinghua University, Beijing 100084, China.

[2]Key Laboratory of Advanced Materials of Ministry of Education of China, Tsinghua University, Beijing 100084, China.

[3]State Key Laboratory of New Ceramics and Fine Processing, Tsinghua University, Beijing 100084, China.

*Corresponding author. Email: **ryu@tsinghua.edu.cn**

[†]These authors contributed equally to this work.

**Abstract:**

**Atomic-scale characterization of spin textures in solids is essential for understanding and tuning properties of magnetic materials and devices. While high-energy electrons are employed for atomic-scale imaging of materials, they are insensitive to the spin textures. In general, the magnetic contribution to the phase of high-energy electron wave is 1000 times weaker than the electrostatic potential. Via accurate phase retrieval through electron ptychography, here we show that the magnetic phase can be separated from the electrostatic one, opening the door to atomic-resolution characterization of spin textures in magnetic materials and spintronic devices.**

**Main Text:**

Imaging structure of matter with ever-higher spatial resolution has been a central theme in science. With the correction of the spherical aberration [1], sub-Angstrom resolution has been routinely available using conventional or scanning transmission electron microscopy (TEM) [2, 3]. Despite a powerful tool for imaging lattice structures with atomic-resolution, transmission electron microscopy encounters large challenges for atomic-resolution magnetic imaging, which is essential for understanding and control magnetism of materials and devices at the atomic scale in spintronics and antiferromagnetic spintronics [4, 5, 6].

The challenge of atomic-resolution magnetic imaging lies in the intrinsically weak interaction between high-energy incident electrons with the magnetic potential in an object. In general, the magnitude of the magnetic interaction is only $10^{-3}$ times (or less) that of the interaction with electrostatic potential [7, 8], leading to very weak magnetic phase compared to the strong electrostatic background. Even worse, the top and bottom surfaces of most (if not all) TEM samples are defected, containing physical damages and/or chemical contaminations of several atomic layers or thicker. The electrostatic potential of the defected surface layers would form large noise to weak magnetic phases. For example, a single atom at the surface of a 100-nm-thick sample would give electrostatic phase stronger than the magnetic phase integrated over the whole thickness.

If the effect of defected surface layers can be neglected, current magnetic imaging methods usually provide a nanometer-scale spatial resolution [9, 10, 11, 12], e.g., the off-axis electron holography in the TEM [13, 14]. Recently, Kohno *et al.* [15] demonstrate real-space visualization of magnetic structure using differential phase contrast (DPC). In order to remove the electrostatic contribution from the total DPC signals, the method uses the so-called "kernel filter" or "B-field filter" that depends on known magnetic structures. The obtained images are then averaged over many unit cells to improve the signal-to-noise ratio, but at the cost of losing atomic resolution. Recently, it is shown that sub-nanometer resolution in magnetic imaging can be obtained by detecting the shift of diffraction patterns using a scanning electron beam[16]. Employing momentum-dependent inelastic scattering in electron energy-loss spectrum, the electron magnetic circular dichroism (EMCD) [17] can provide element-selective magnetic moments plane by plane [18]. Other EMCD-based schemes have also been proposed to detect magnetic signals [19], yet to be realized experimentally. Since the cross-section of core-loss inelastic scattering is less than that of elastic scattering by a factor of several orders of magnitude [20], high electron dose would be required for acceptable signal-to-noise ratio. A common problem with known methods for magnetic imaging is the lack of a convenient mechanism to eliminate the contribution of defected surface layers, resulted in limited precision and accuracy in experimental measurements.

Using multislice ptychographic phase retrieval at deep-sub-angstrom resolution, here we show that the magnetic contribution to the phase of the electron wave can be accurately measured and the effect of defected surface layers can be eliminated, resulting in atomic-resolution imaging of magnetic structures.

We use antiferromagnetic hematite ($\alpha$-Fe$_2$O$_3$) as a model system. The $\alpha$-Fe$_2$O$_3$ phase has a high Néel temperature $T_N$ at about 950 K [21]. As shown in **Fig. 1**a, the magnetic moments of the iron ions form ferromagnetic bilayers between hexagonal-close-packed (0001) planes of oxygen. The iron bilayers are coupled antiferromagnetically across the oxygen layers. Note that the magnetic field in the objective lens of a TEM is about 2 T, well below the spin-flop field (order of 10 T) of α-Fe$_2$O$_3$ [22].

When a high-energy electron beam transmits through a thin sample, the phase shift induced by the electrostatic and magnetic potentials is [23]:

$$\Delta\varphi = \sigma V_z - \frac{2\pi e}{h} A_z = \varphi_V + \varphi_A \quad (1)$$

where $\sigma = 2\pi e/h\nu$ is the interaction constant, $V_z$ the projected electric potential, $A_z$ the path integral of magnetic vector potential along the electron beam direction described in Methods, $h$ the Planck's constant. The two terms on the right side describe the electrostatic and magnetic phase, respectively.

The electrostatic potential and spin density for computing magnetic vector potential were obtained by density functional theory (DFT) calculations using the full-potential linearized augmented plane wave (FP-LAPW) method [24]. The electrostatic, magnetic and total phase are derived (see Methods for details) and shown in **Fig. 1**, along with their diffractograms. The electrostatic phase (**Fig. 1**c) is by far the largest contribution to the total phase (**Fig. 1**b). Different from the electrostatic phase that is peaked at atoms, the magnetic phase (**Fig. 1**d) is distributed in between magnetic atoms. For convenience, the electrostatic and magnetic phases are distinguished as "on-site" and "off-site" phase, respectively. When averaged in the (0001) plane, the magnetic phase has an antiferromagnetic undulation in the [0001] direction, with the undulation amplitude of 0.17 mrad/nm. Note that the peak in the on-site electrostatic phase is 790 mrad/nm, but the peak in the off-site magnetic phase is only 0.47 mrad/nm, three orders of magnitude smaller than the electrostatic one. The ratio of the magnetic to total phases is plotted in **Extended Data Fig. S2**, indicating that very high accuracy in phase measurement is required to separate the magnetic phase from the electrostatic one.

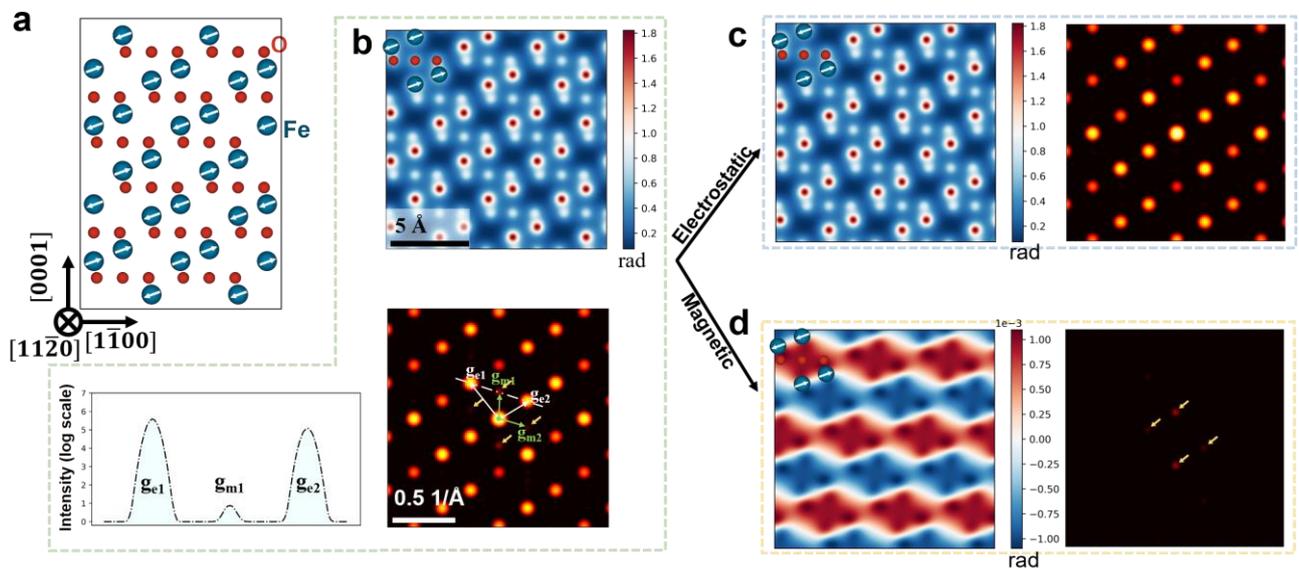

**Fig. 1. The contributions of electrostatic and magnetic potentials to the phase of the incident electron wave. The potentials are calculated using density-functional theory. a**, Spin structure of α-Fe$_2$O$_3$ at room temperature; **b**, The total phase, it's diffractogram, and the intensity profile along the white dash-dotted line in the diffractogram; **c**, Electrostatic phase and it's diffractogram; **d**, Magnetic phase and it's diffractogram. The sample thickness in the simulations is 2.3 nm. The diffractogram intensities are displayed on a logarithmic scale to show the weak magnetic

reflections.

In **Fig. 1**, the diffractograms of the total, electrostatic, and magnetic phase are also shown. The electrostatic phase contributes to the strong reflections, defined by the reciprocal lattice $g_e(h,k) = h \times g_{e1}/2 + k \times g_{e2}/2$, with the basis vectors $g_{e1} = (\bar{1}104)$ and $g_{e2} = (1\bar{1}02)$. The magnetic phase contributes to the weak reflections, the positions of which are defined by $g_m(h,k) = h \times g_{e1}/2 + k \times g_{e2}/2$ (both h and k are odd integers), as indicated by the arrows in **Fig. 1d**. The intensity of the magnetic reflection (0003) is about $2.0 \times 10^{-5}$ times that of the electrostatic reflection $(\bar{1}104)$. Because the mean value of the magnetic phase is zero, the magnetic phase has no contribution to the reciprocal origin (0,0). As a result, the electrostatic and magnetic phases are well separated in the reciprocal space. The strong reflections [h and k being even integers, including the origin (0,0)] are contributed by the on-site electrostatic potential, and the weak reflections (h and k being odd integers) by the off-site magnetic potential. Applying masks in reciprocal space (shown in **Extended Data Fig. S3c**) and performing Fourier transform, the electrostatic and magnetic phases are separated in real space. Hereafter we name the process to extract weak magnetic phase from the total phase as "Fourier separation". The challenge is to obtain the total phase of sufficient accuracy, since the magnetic phase is more than 1000 times weaker than the total phase.

We use the multislice electron ptychography with adaptive propagator to measure the phase of the object. Ptychography is a phase-retrieval method using coherent diffraction imaging in the scanning mode [25, 26, 27, 28]. As an electron probe scans over a two-dimensional grid of an object, two-dimensional coherent diffraction patterns are collected using a pixelated array detector with high dynamic range [29], as schematically shown in **Fig. 2a**, forming the so-called 4D-STEM dataset (2D scanning in real space and 2D diffraction in reciprocal space). Various algorithms have been proposed to recover the phase of the object encoded in the datasets [30, 31, 32, 33, 34, 35]. The experimental challenges like residual lens aberrations, dynamic electron diffraction [36], sample drift during data collection [37], and the misorientation between crystal zone axis and electron beam [38] can be corrected, providing accurate phase of the object. Ptychography also offer a higher signal-to-noise ratio than conventional STEM imaging modes at the same level of electron dose [39, 40].

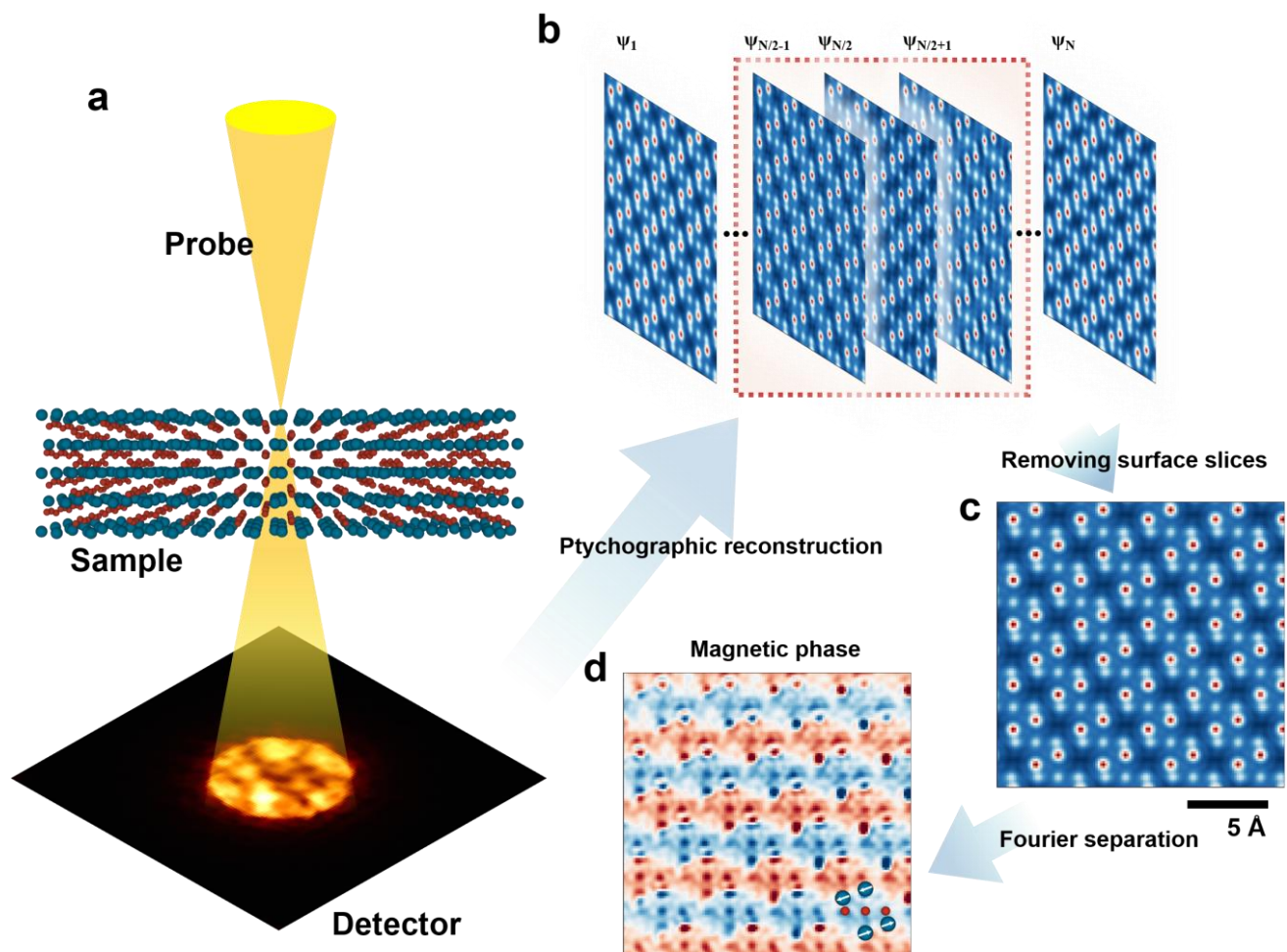

**Fig. 2. Workflow of extracting magnetic phase from total phase. a**, Experiment set up for electron ptychography; **b**, Reconstructed phase images of multiple slices; **c**, The averaged phase image of the central slices; **d**, The magnetic phase extracted by Fourier separation.

The workflow and typical results are schematically shown in **Fig. 2**. In order to differentiate the defected surface layers from the pristine bulk, multislice reconstruction is necessary, as shown in **Fig. 2**b. With the contributions from the surface layers removed, the phases of the central slices represent the pristine bulk, as shown in **Fig. 2**c. In the last step, the Fourier separation is applied to the total phase to obtain the electrostatic phases (not shown) and the magnetic phase (**Fig. 2**d), which is undulating in the [0001] direction.

Because it is necessary to measure the phases very accurately, we performed simulation tests on the sensitivity of above method on the electron dose, which influences the accuracy of experimental measurements due to the shot noise. The doses of $10^3$ e/Å$^2$ ~ $10^6$ e/Å$^2$ are considered. As reference, a typical high-angle annular dark field (HAADF) or DPC image is recorded with a dose of $10^4$ e/Å$^2$ ~ $10^6$ e/Å$^2$. The results are shown in **Fig. 3** and **Extended Data Fig. S4**. The total phases of the central slices reconstructed by ptychography are shown in **Fig. 3**a. High-quality total phases can be obtained for an electron dose down to $4.5 \times 10^3$ e/Å$^2$, with all the atomic columns including oxygen are clearly imaged, although single diffraction patterns are quite noisy at this dose, as shown in **Extended Data Fig. S5**.

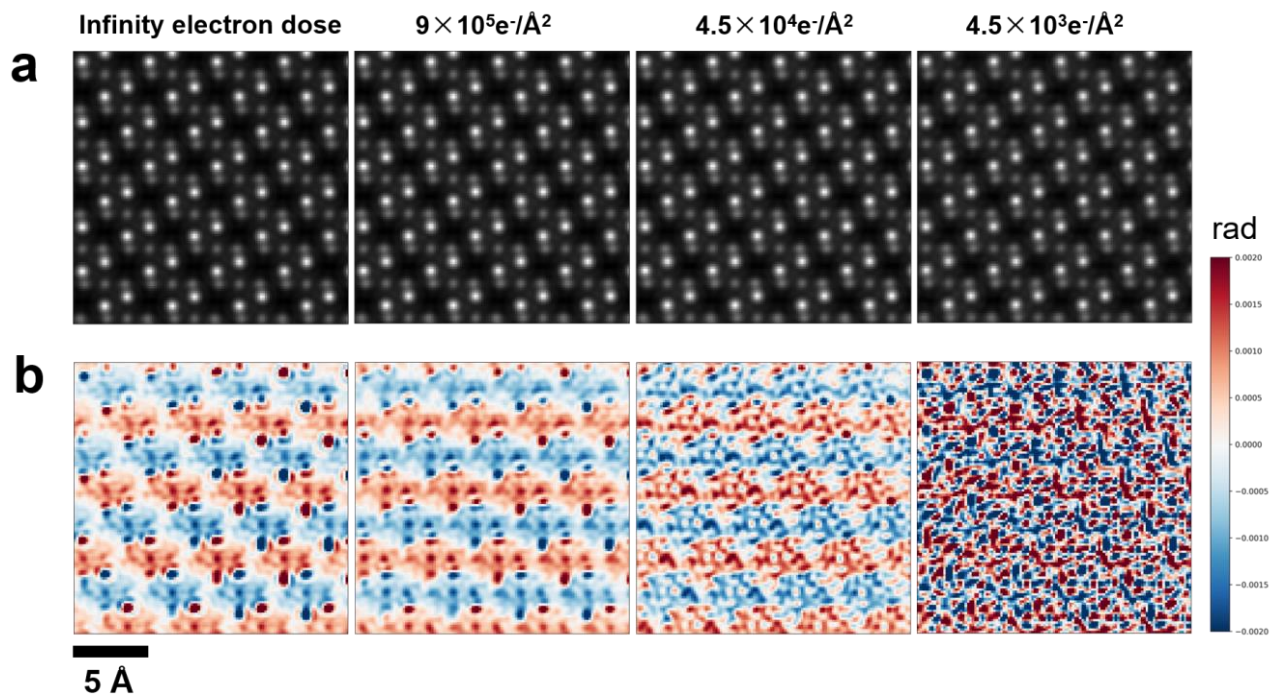

**Fig. 3. The total and magnetic phases obtained with simulation datasets for different electron doses. a**, the total phases via ptychographic reconstruction, **b**, the magnetic phases via Fourier separation. The sample thickness is 2.0 nm along, [11$\bar{2}$0] zone axis.

The magnetic phases extracted by Fourier separation are shown in **Fig. 3**b. Since the magnetic phase is extremely weak, much higher electron dose is required to obtain a reasonable signal-to-noise. As shown in **Extended Data Fig. S3**b and **3**c, the magnetic phase becomes very noisy at the dose of $4.5\times10^4$ e/Å$^2$, and almost indiscernible at a dose of $4.5\times10^3$ e/Å$^2$. It should be noted that, in experimental datasets, there are other sources of noise that are not considered in the simulations shown in **Fig. 3**, including point defects, dislocations, strain field, etc. Therefore, the simulations like **Fig. 3** only suggest a rough estimate of the lower bound of the electron dose for detecting magnetic phase. In experiments, higher electron dose may be needed.

The simulation results show another challenge for the detection of weak magnetic phase, i.e., the disturbance of the strong electrostatic phase. Due to numerical errors inevitable in the reconstruction process, the trace of electrostatic phase appears in the magnetic phase at the atomic columns, even at the infinite electron dose. Fortunately, the electrostatic and magnetic phases are separated in both real and reciprocal spaces, being on-site and off-site, respectively (**Fig. 1**c and **1**d). It means that the on-site phase appeared in the magnetic phase image can be attributed to the trace of the electrostatic phase, and its influence on the interpretation of the magnetic phase is limited.

**Fig. 4** shows the experimental imaging of the total and magnetic phases. The electron dose we used in experiment is $9.0\times10^5$ e/Å$^2$. In the diffractograms, the magnetic reflections can be clearly identified. The intensity of the magnetic reflection (0003) is $7.9\times10^{-5}$ times that of the electrostatic reflection ($\bar{1}$104). The intensity ratio is the same order of magnitude as the theoretical prediction (~$2.0\times10^{-5}$, **Fig. 1**b). Possible sources of the discrepancy include numerical errors for the measurements of such weak signals, the temperature effect, diffuse scattering due to possible point defects, and the uncertainties in the DFT calculations of a system of strongly-correlated electrons.

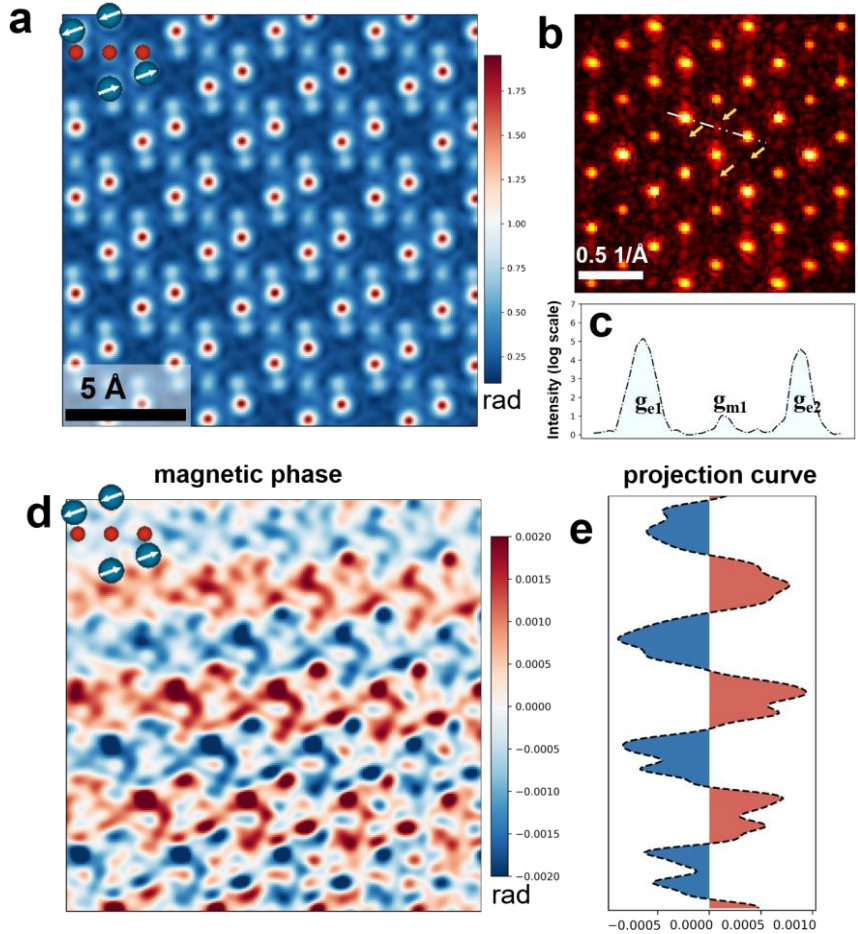

**Fig. 4. Experimental magnetic phase imaging of α-Fe$_2$O$_3$. a**, The total phase averaged over the central slices and **b** the corresponding diffractogram (on a logarithmic scale). The arrows indicate the magnetic reflections of the lowest order. **c**, The intensity profile along the line connecting g$_{e1}$, g$_{m1}$, and g$_{e2}$, as indicated in the diffractogram in **b** by the white dash-dotted line. **d**, The magnetic phase extracted from the total phase by Fourier separation. The Fe ions are overlaid on the upper left corner with arrows indicating their spin directions. **e**, The projection of the magnetic phase to the vertical axis. All the phases correspond to a slice of 2.3 nm thick. The electron dose is $9.0\times10^5$ e/Å$^2$.

**Fig. 4**d and e shows the magnetic phase extracted from the total phase via Fourier separation. The Fe ions are overlaid in the upper left corner, indicating the antiferromagnetism along the [0001] direction. The magnetic phase averaged in the (0001) plane is shown on the right of **Fig. 4**d, clearly revealing the phase undulation in the [0001] direction. The measured undulation amplitude is 0.40 mrad/nm, consistent with the theoretical prediction (0.17 mrad/nm) considering numerical errors in the measurements of weak signals. As expected from the simulations shown in **Fig. 3**, the trace of on-site electrostatic phase also appears in the magnetic phase image due to numerical errors.

It should be noted that magnetic reflections occur only in antiferromagnetic materials. For ferromagnetic materials, there is no sharp magnetic reflections but diffuse scattering at magnetic domain boundaries or other magnetic microstructures. Applying Fourier separation to the diffuse scattering would require even higher signal-to-noise ratio in the total phase and in-turn higher electron dose. Another condition for atomic-resolution magnetic imaging of ferromagnetic materials is a sample environment free of magnetic field, otherwise magnetic textures would be washed out easily by the magnetic field in a TEM. High-resolution field-free microscopes have appeared [41].

In summary, atomic-resolution imaging of magnetism has been realized through phase retrieval of multislice electron ptychography with adaptive-propagator. The phase is accurately measured, enabling direct separation of weak magnetic phase (< 1 mrad/nm) from strong electrostatic phase (~1 rad/nm) in both real and reciprocal spaces. The method would find wide applications first in antiferromagnetic materials/spintronics, then in ferromagnetic ones when high-resolution field-free microscopes are widely available.

**Methods:**

Sample preparation

The α-Fe$_2$O$_3$ single crystal was purchased from Prmat (Shanghai) Technology Co., Ltd. Cross-section TEM samples were prepared using a focused ion beam (FIB) instrument. The samples were thinned down to 20 nm using an accelerating voltage of 30 kV with a decreasing current from 240 pA to 50 pA, followed by fine polishing with an accelerating voltage of 5 kV with a draft of 20 pA. The sample thickness measured by EELS log-ratio method [42] is 20 nm, agree well with sample thickness obtained from multislice ptychography.

Ptychography experiments and reconstructions

The scanning diffraction datasets were acquired using a pixel-array detector (EMPAD) equipped on a probe aberration corrected FEI Tian Cubed Themis G2 operated at 300 kV. The convergence semi-angle was set to 25 mrad. Each diffraction pattern has a dimension of 128×128 and the nominal camera length is 285 mm, giving a reciprocal pixel size of 0.055 Å$^{-1}$ (1.1 mrad). Regular scanning grid was used with 128×128 scanning positions. In-focus condition combined with 0.367 Å scan step size was adopted. The beam current is 20 pA and the dwell time is 1 ms. For the results shown in **Fig. 2** the diffraction patterns were padded to 200×200 to get a real space pixel size of 0.091 Å. Ten object slices were used for adaptive-propagator ptychography reconstructions. Drift correction and mixed-state algorithm with six probe modes were used.

Density functional theory calculations

The collinearly spin polarized DFT calculations were carried out using the full-potential linearized augmented plane wave (FP-LAPW) [24] as implemented by WIEN2k. We adopted the minimal magnetic supercell which only contains two Fe atoms with different spin directions. To correctly consider the strongly correlated $3d$ orbitals of Fe, PBE + $U$ approach was adopted ($U$ = 2.5 eV). The After convergence, electrostatic potential and spin density were chosen as output for the next multislice simulation.

Multislice simulation of 4D datasets

An electron microscopy simulation program based on details referred by Kirkland [43] was used to simulated 4D datasets for ptychographic reconstruction. Instead of independent atom model (IAM) potential, both DFT electrostatic potential and magnetic potential corresponding the magnetic structure are chosen as potential input. The 4D datasets are generated for the high tension of 300 kV, the convergence semi-angle of 25 mrad and in-focus condition as same as experiments.

Calculation of phase shift of magnetic vector potential

The magnetic vector potential is given by [8]:

$$\boldsymbol{A}(\boldsymbol{r}) = \frac{\mu_0}{4\pi} \frac{\boldsymbol{\mu} \times \boldsymbol{r}}{r^3} * M(\boldsymbol{r}) \tag{1}$$

where $\mu_0$ is the vacuum permeability, $\boldsymbol{\mu}$ the unit vector parallel to dipole moment, * denotes a convolution, the magnetic moment $M(r) = \mu_B \rho(r)$, $\mu_B$ is the magnetic moment of one electron, *i.e.*, Bohr magneton, $\rho(r) = \rho_{up}(r) - \rho_{down}(r)$ is electron spin density, $\rho_{up}(r)$ and $\rho_{down}(r)$ are the electron charge density of spin-up and spin-down, respectively. $\rho_{up}(r)$ and $\rho_{down}(r)$ are obtained with DFT calculations.

Like the projected potential used in electron microscopy image simulations, magnetic vector potential is integrated along probe incident direction $z$ of a certain thickness. The phase shift $\varphi_A$ caused by magnetic vector potential is then:

$$\varphi_A = -\frac{2\pi e}{h} \int \boldsymbol{A}(\boldsymbol{r}) dz \qquad (2)$$

**Acknowledgments:** In this work we used the resources of the Physical Sciences Center and Center of High-Performance Computing, Tsinghua University.

**Funding:** This work was supported by the National Natural Science Foundation of China (51788104 and 51525102).

**Author contributions:** R.Y. designed and supervised the research. J.C., H.S., and W.Y. performed simulations, experiments, and reconstructions. All authors performed data analysis, discussed the results, and co-wrote the manuscript.

**Competing interests:** The authors declare no competing financial interests.

**Data and materials availability:** All data used to evaluate the conclusions of the paper are in the paper and/or Supplementary Materials. Additional data related to this paper are available from R.Y.

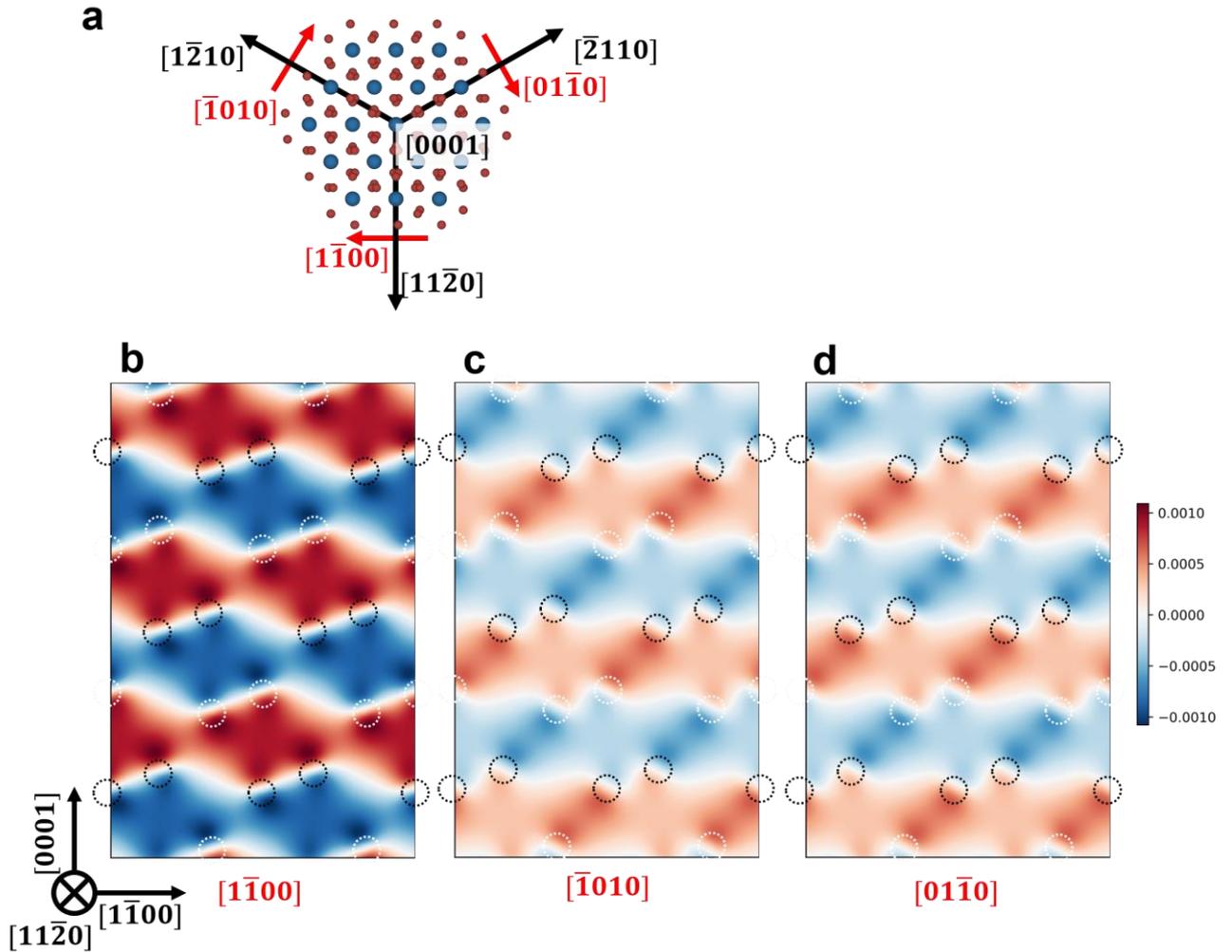

**Extended Data Fig. S1**

**Projected magnetic phase along the $[11\bar{2}0]$ zone axis with different spin orientations. a**, The schematics of three spin orientations. The arrows show the $[11\bar{2}0]$, $[1\bar{2}10]$, and $[\bar{2}110]$ directions, and the spin orientations approximately lying in the $[1\bar{1}00]$, $[\bar{1}010]$ and $[01\bar{1}0]$ directions. For a given $[11\bar{2}0]$ zone axis, the projected magnetic phase for the three spin orientations are shown in **b, c,** and **d**, respectively. In this study, we consider the projected magnetic phase with the spin orientation in $[1\bar{1}00]$, which is twice of those in the other two spin orientations. The black and white circles denote atoms of opposite spin direction. In the simulations, the sample thickness is 2.3 nm, and all the three spin orientations have a 20° tilt away from the basal plane.

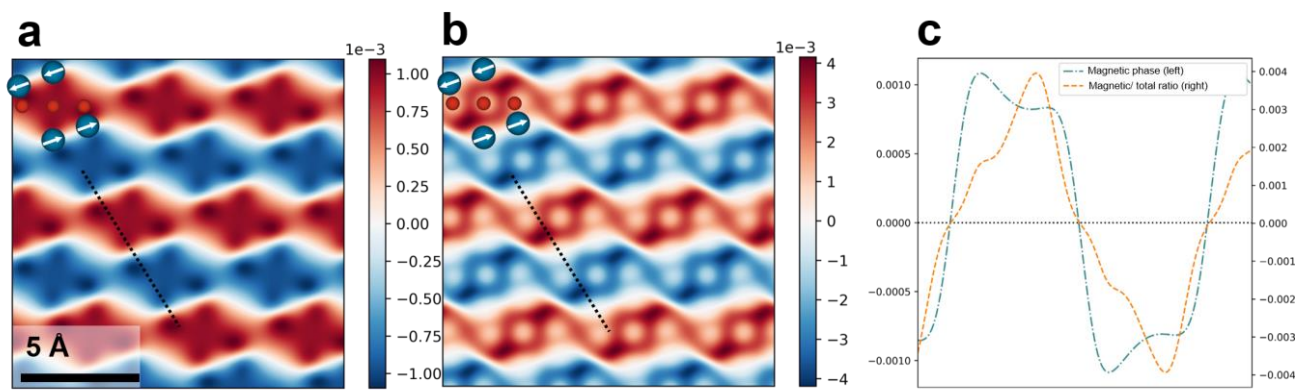

**Extended Data Fig. S2**

**a**, Magnetic phase of a sample of 2.3 nm thick; **b** Ratio of the magnetic phase to the total phase; **c**, The profiles of the magnetic phase and the magnetic/total ratio along the line indicated in **b**.

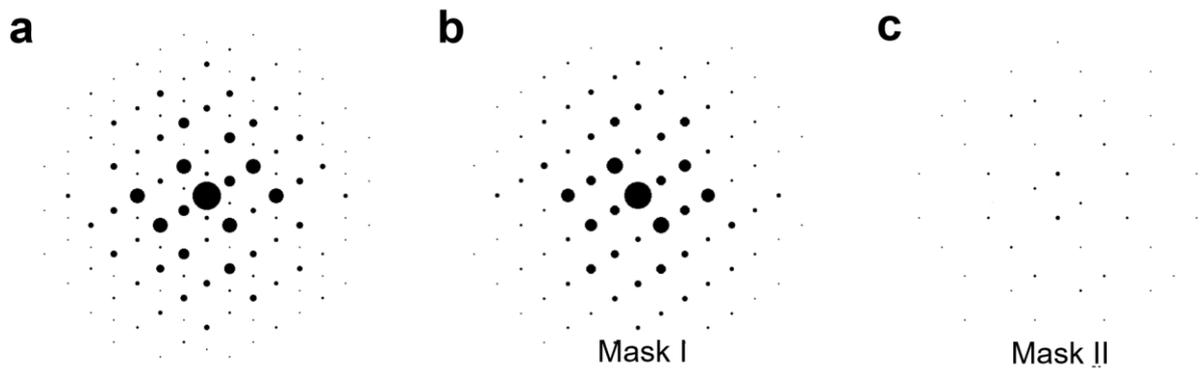

**Extended Data Fig. S3**

**a**, Simulated diffraction pattern of α-$Fe_2O_3$ with magnetism considered, in the $[11\bar{2}0]$ zone axis. **b**, Simulated diffraction pattern of α-$Fe_2O_3$ without magnetism considered, in the $[11\bar{2}0]$ zone axis. It is same as Mask I covering the principal reflections in **a**, which are contributed by the electrostatic potential $\varphi_V$. **c** Mask II covering extra reflections in **a**, which are contributed by the magnetic potential.

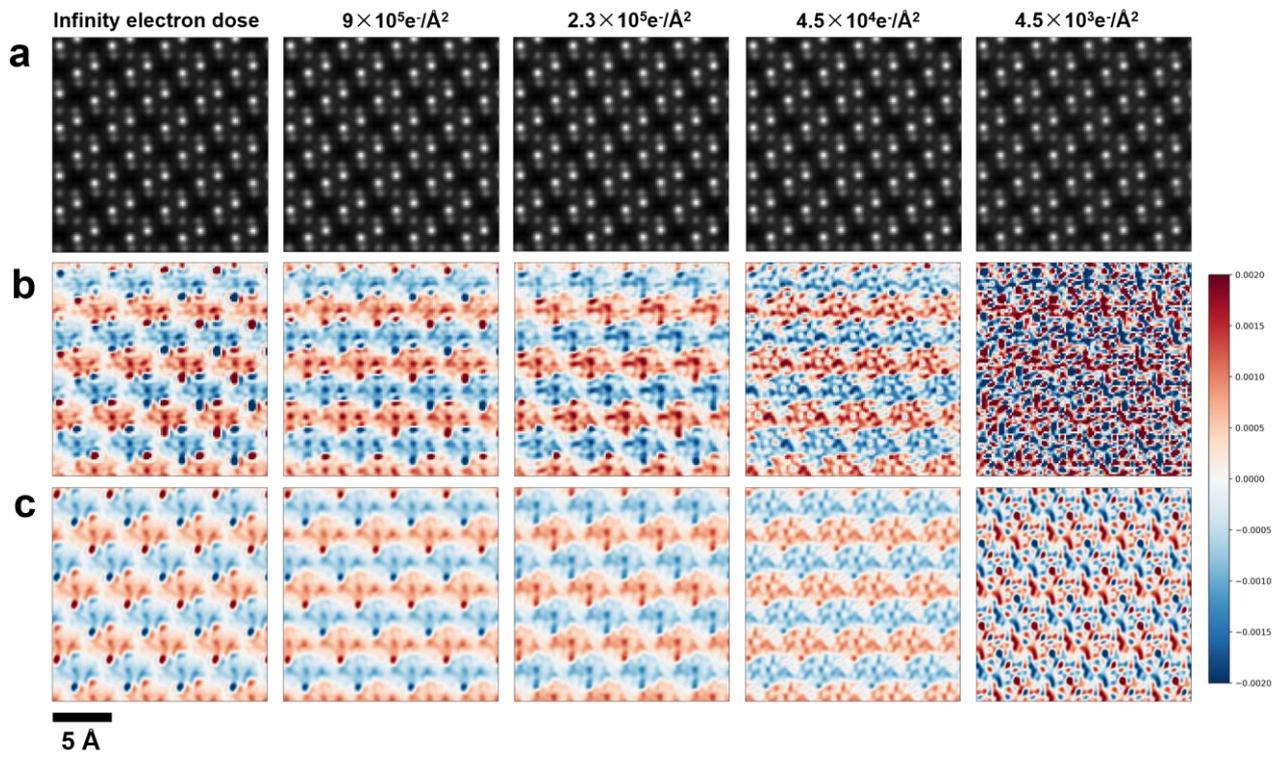

**Extended Data Fig. S4**

**Simulation test of the total and magnetic phases at different electron doses.** The columns from left to right show **a** the total phases via ptychographic reconstruction, **b** the magnetic phases via Fourier separation, **c** the magnetic phases via Fourier separation and subsequent unit-cell-averaging, and tiling. The sample thickness is 2.0 nm along, [11$\bar{2}$0] zone axis.

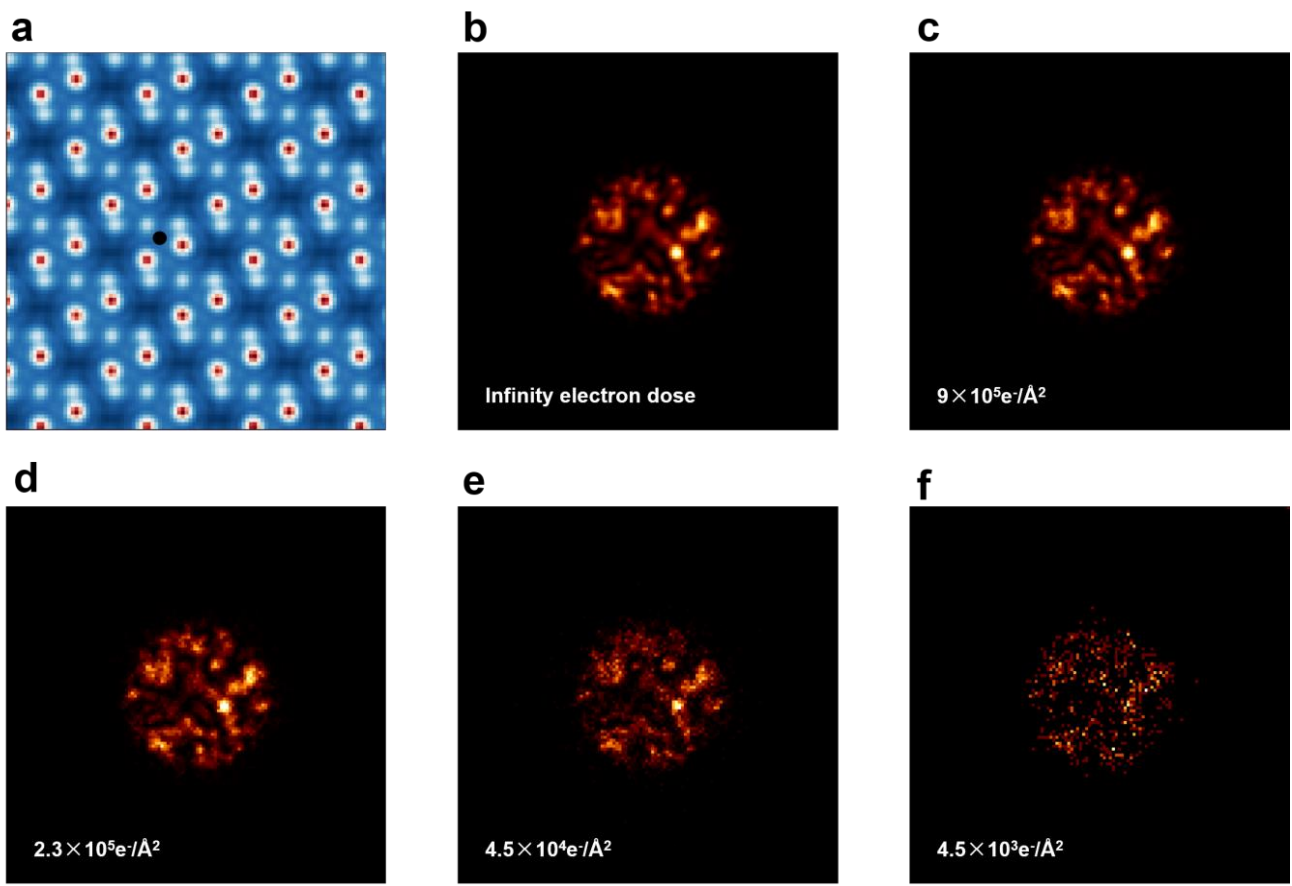

**Extended Data Fig. S5**

Comparison between CBED at different electron doses. **a**, Phase reconstructed by ptychography, the black point indicating the probe position for the Diffraction patterns. **b,c,d,e,f**, Single Diffraction patterns at different electron doses.